# Performance Evaluation of Deep Convolutional Maxout Neural Network in Speech Recognition

Arash Dehghani
Department of Biomedical Engineering
Amirkabir University of Technology
Tehran, Iran
arash.dehghani@aut.ac.ir

Seyyed Ali Seyyedsalehi
Department of Biomedical Engineering
Amirkabir University of Technology
Tehran, Iran
ssalehi@aut.ac.ir

*Abstract*— In this paper, various structures and methods of Deep Artificial Neural Networks (DNN) will be evaluated and compared for the purpose of continuous Persian speech recognition. One of the first models of neural networks used in speech recognition applications were fully connected Neural Networks (FCNNs) and, consequently, Deep Neural Networks (DNNs). Although these models have better performance compared to GMM / HMM models, they do not have the proper structure to model local speech information. Convolutional Neural Network (CNN) is a good option for modeling the local structure of biological signals, including speech signals. Another issue that Deep Artificial Neural Networks face it, is the convergence of networks on training data. The main inhibitor of convergence is the presence of local minima in the process of training. Deep Neural Network Pre-training methods, despite a large amount of computing, are powerful tools for crossing the local minima. But the use of appropriate neuronal models in the network structure seems to be a better solution to this problem. The Rectified Linear Unit neuronal model and the Maxout model are the most suitable neuronal models presented to this date. Several experiments were carried out to evaluate the performance of the methods and structures mentioned. After verifying the proper functioning of these methods, a combination of all models was implemented on the FARSDAT speech database for continuous speech recognition. The results obtained from the experiments show that the combined model (CMDNN) improves the performance of ANNs in speech recognition versus the pre-trained fully connected NNs with sigmoid neurons by about 3%.

I. INTRODUCTION

Decades ago, researchers obtained successful results from ANNs with a hidden layer of nonlinear neurons to predict HMM states from acoustic coefficients. At that time, though, hardware equipment, as well as learning algorithms, were inadequate to train neural networks with a large number of hidden layers and high volume of data. For this reason, a negligible increase in the performance of using neural networks compared to the GMM method was not so significant or, in other words, not cost-effective. Over the past few years, advances in machine learning algorithms, as well as the development of computer hardware, have led to the emergence of better methods for training ANNs, including a large number of hidden layers with nonlinear neurons [12]. In recent years, another method has been used to determine the relationship between the HMM states and input frames, which uses an ANN instead of the GMM model. The input of this neural network is the speech frame acoustic signal frames and the output produces the posterior probabilities required by the HMM model [1]. However, fully connected Neural Networks, despite their simplicity, have weaknesses. First, they will not be able to model the local information of the signal. Second, neurons with a sigmoid operator function may degrade the gradient during training. This phenomenon happens, because when the input of the sigmoid function is not near zero, the gradient of the function tends to be zero and practically does not modify the weights for the lower layers [2]. In addition, because of the greater influence and co-adaptation that network parameters have on each other, the power of its generalization to the test data is reduced and the divergence phenomenon occurs. Recently, some methods have been developed to overcome these problems. For the purpose of refining the network against small changes in the signal, Time Delay Neural Networks[3] (TDNNs) and Recurrent Neural Networks [4](RNNs) have been introduced. Consequent to this, the developed and optimized model of TDNNs, called Convolutional Neural Network (CNN) [5], was introduced by Yann Lecun, which largely addresses the problem of modeling local signal information. However, pre-training methods have been introduced to address the gradient degradation problem during network training that causes localized minima [6][7]. Although they have a large number of computations, by using these methods, DNNs can be transmitted from local minima to an optimum point. Recently, new neuronal models have been developed that effectively eliminate the problem of local minimums. The Rectified Linear Unit (ReLU) [8]and its generalized Maxout neuron [9]are among the most important of these models. Another issue that we face after passing deep neural networks from the local minima and converging them into the training data, is the power of their generalization to the test data. To solve this problem, the Dropout training method is used in the network training process. The details and method of operation of all methods and structures will be described in the coming section.

II. METHODS, STRUCTURES, AND PRIOR WORKS

In this section, we will try to explore new methods and solutions that increase the efficiency and accuracy of Artificial Neural Networks, separately in detail.

*A. Pre-training*

Due to a large number of local minima, deep neural networks will not usually converge [6]. However, with the initialization of the network weights, many of the local minima can be avoided. Pre-training methods are used to find the initial values of network weights. Fundamentally, pre-training methods are used with the aim of releasing the learning process from the existing local minima along the way, as an essential barrier to the training process. These methods seek to find a good starting point for network weights, and in addition to facilitating the network learning process, improve the power of network generalization [10].

There are some methods to Pre-train neural networks. In 2006, Hinton presented a technique based on Restricted Boltzmann Machines (RBM) for pre-teaching multi-layer neural networks in order to reduce nonlinear dimensionality [7]. In this method, the multi-layer network is broken up to the corresponding RBMs, and pre-training is provided through these RBMs. In 2015, Seyyed Salehi presented a layer-by-layer pre-tutoring method for pre-training deep associative networks for the purpose of extracting essential components [6]. But the bidirectional version of this method is provided for the pre-training of deep neural networks with another association application [10]. This method is used as a new way to train another association of deep neural networks, which is used to converge fully connected networks with sigmoid and sigmoid-tangent nonlinear neurons. In this work, we use this method to pre-train the neural network with sigmoid nonlinearity.

*B. Convolutional Neural Networks*

The most important drawback of fully connected neural networks for their applications in image and speech processing is that they do not have the mechanism to deal with converter changes or distortion of input data. In principle, a fully connected network with enough capacity can be configured and trained to produce unchanged outputs with these changes. Doing this would probably result in the creation of multiple neurons with distinct weight patterns that have been placed in different places in the input space [11]. Convolutional Neural Networks carry local extraction by limiting the input field of the hidden neurons and forcing them to be local. In other words, in CNNs, spatial unchanging will automatically be accomplished by the obligatory repetition of the configuration of weights in space [11]. CNNs combine three structural ideas to achieve spatial irregularity and distortion. These are local receiver units, weight sharing and Spatial and temporal sampling (pooling), and integration. With local receptive regions, neurons can detect and extract basic visual features such as edges, endpoints, corners in a picture, or local features in the speech spectrum. These features will eventually be combined in the upper layers [11]. Yan Lecun introduced CNNs in 1995 and evaluated their performance in image and sound processing applications [11][5]. This structure of neural networks has been used in many applications. After CNNs showed high ability to obtain immutability with location displacement in the category of image processing [12], Honglak Lee et al. in 2009 used them as unsupervised learning for speech recognition and showed that they had good performance in speech recognition[13]. In 2012, Abdel_Hamid et al. assume that spatial variations along the time dimension can be modeled by the dynamic properties of HMM states, used one-dimensional CNNs to model local variations along the frequency side[14]. In their network structure, local filters and pooling layers are applied along the frequency side. By using this technique, the speaker's variations can be modeled and it leads to achieving a better recognition performance. A few years later, Abdel-Hamid et al. expanded their model and compared the different pooling methods and they concluded that maximum selection in the pooling layer has better performance compared with average operation [15]. Sainath et al. also did the same for comparing weight-sharing methods as well as different methods of pooling strategies, and they yielded results similar to those of Abdul Hamid [16].

*C. Appropriate Neuronal Models*

In this section, the neuronal models proposed to improve the performance of Artificial Neural Networks are reviewed.

*1) Rectified Linear Unit (ReLU)*

Glorot et al. (2011), based on the biological neuronal model presented by Dayan and Abbott in 2001 [17], showed that using a linear neuronal model that has been rectified, in ANNs instead of hyperbolic tangent or sigmoidal neurotransmitter models, would improve their performance [8]. Despite its hard nonlinearity and its inflexibility in zero points, it is biologically similar to natural neurons and improves the performance of ANNs and their training process. Its approximate equation is shown in relation (1).

$$(1)$$

After verifying the quality of the Rectified Linear Unit neuronal function by Glorot, Zeiler et al. in 2013, they used this neuronal model for speech recognition and obtained good results [18]. The ReLU neuronal model was used in many speech recognition applications and performed better results compared to previous models [19][20][21][22]. This neuron model has better performance without using bias (biases equal to zero) [18]. Also due to the instability of its linear part in the network, weight normalization and sometimes limiting the output of the network layers is required [21]. A ReLU neuronal model, like biologic neurons, causes sparsity in the network. Experiments show that the process of network training will be improved when the neuron is off or linear. It may seem that the process of training is difficult due to the saturation of this neuron at zero, which is why the developed model of the neuron was named Soft-Plus. This model has a more nonlinear softness than the original model [8].

*2) Maxout*

As previously mentioned, the ReLU neuronal model suffers from saturation at zero and divergence in its linear region. However, improved structures such as Soft-Plus could slightly deal with these problems, but such models did not completely get rid of these issues and were constantly subject to saturation and divergence. In 2013, Goodfellow et al. presented a model called Maxout. Despite its simplicity, this model largely fixed the defects of the ReLU model. Maxout, by fixing saturation mode, provides better network training and easier convergence. The function of this neuron is in a way that, unlike the linear neurons and the sigmoid, it always crosses the gradient and it does not vanish the propagated gradient. This property is due to the fact that its output, at any time, is equal to the output of a neuron with a linear operator function (y = x), which has a maximum value relative to a group of neurons, as can be seen in equation (2). So the derivative is always equal to one. Also, the process of maximization is considered a feature selector [23]. Typical structure of Maxout networks is shown in Fig. 1. For example, a 4 hidden layer network is displayed. The Maxout structure consists of a fully connected and one-dimensional maxpooling layer (used in CNNs). The maxpooling layer is then connected to the fully connected layer to implement the Maxout structure in a total of two layers by applying the localization between the outputs of the previous neurons. In general, a fully connected layer and a pooling layer form a maxout layer.

$$h_i(x) = \max(z_{ij}) \quad i \in [1,k] \tag{2}$$

$$z_{ij} = x^T W_{\ldots ij} + b_{ij}, \text{ and } W \in R^d$$

In 2013, after presenting the Maxout Model by Goodfellow et al., Kai and colleagues, as well as Miao et al., used this model for speech recognition applications the same year. Each of these two groups introduced themselves as the first to use this structure for the speech processing category [24][25]. Kai et al. performed experiments on DNNs including Sigmoid, ReLU, and Maxout, which showed that the Maxout model had the best performance compared to other neuronal models [25]. These researchers, one year later in 2014, expanded their neural network structure and used the Maxout model in a CNN structure and achieved good results [23]. Subsequently, various research groups benefited from the Maxout model in the structure of their acoustic model for speech recognition and obtained better results than previous structures [26][27][28].

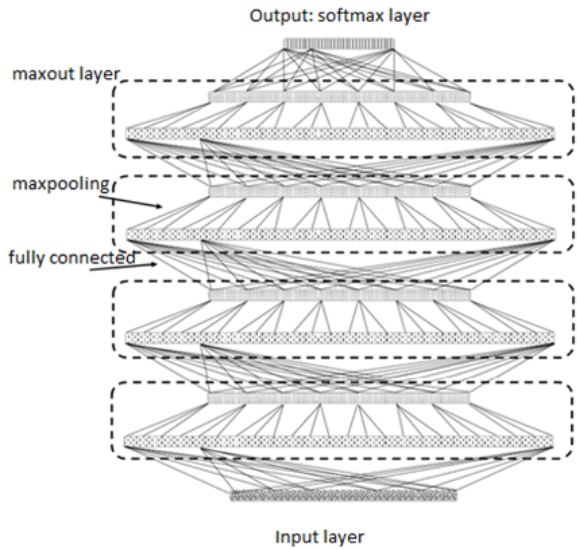

Figure 1. Typical structure of fully connected neurons with Maxout implemented on FARSDAT.

### D. Dropout

Deep neural networks, with nonlinear functions in different layers, can learn very complex relationships between inputs and outputs. However, with relatively limited volumes of data, such learning of complex interpersonal relationships can be problematic and may interfere with the ability of the network to generalize to non-trained content. As a result, these complex relationships between the training data will remain in the same phase and will not be able to be generalized to the test data. This phenomenon will lead to the overfitting problem [29]. Many ways are suggested to solve this problem, each of which has its own characteristics. It can be said that if there is no problem in terms of calculation, time, and cost, then using plenty of DNNs with the structure appropriate to the testing and using their results to predict the final results of the test data seems appropriate and acceptable (Bagging method). This method, in most cases, improves the performance of the final model, but this method will be more efficient if the structures of the models are different. But finding the optimal point for each model with different structures is very difficult and comes with a lot of calculations. In addition, the training of different models and with different data requires a large amount of data, and it may not be easy to provide a massive amount of data. Dropout is a method that provides solutions for both of the above issues. Srivastava et al. presented Dropout in 2013. This method provides a way to combine different models, with different structures, thereby preventing the overfitting of the networks over testing data. The name Dropout was transmitted from the drop (deleting from the network) units [29]. The process of eliminating neurons is random. Each neuron will have a P-value relative to other neurons in the network. This number is between zero and one when it is closer to one, it creates a greater probability that the neuron will be present in the network structure, and the closer it gets to zero, the probability of that neuron in the network is less and definitely, its effect on network performance will be decreased. When testing, the averaging of the results of all tiny networks cannot be easily accomplished but is done by the Dropout method. The process is to say that the output weights (or the output of the neurons) of the neurons that have been present in the network with a probability of P during network training, will be multiplied by P according to the equation (3) during the test. This process makes the output of each neuron more consistent with reality when testing the network. By performing this averaging, it is observed that the performance of the network is improved at the time of testing, and overfitting is prevented [29].

$$W_{test}^{(l)} = pW^{(l)} \tag{3}$$

### E. Network Sustainability

As long as limited-function operators such as sigmoid functions are used, the weights and output of the neurons are always limited and will not reach infinity. But when we use neurons that are not limited by output and are capable of generating very large numbers, the risk of instability and numbing numbers towards infinity will threaten the network at any time. In order to protect the network against instability, we must limit the weights and output of network neurons in some way. The Weight Normalization method is appropriate for this purpose [30]. In this method, the magnitude of the vector of network weights is limited to a fixed number such as C, and it does not allow extending of this number. In other words, according to the relation $\|W\| < C$, the magnitude of the weights are enclosed in a sphere with a radius of C which is referred to as maximum rate in this way. The magnitude of the vector of weights W is obtained in accordance with equation (4).

$$\tag{4}$$

Which i specifies the number of elements of the vector W and e denotes the numerical value of each element. As long as the magnitude of the weight W is less than constant C, no action will be taken on the weight vector. But when the size W raises from C, the weight vector values are corrected so that the magnitude of it will be equal to C. The benefit of this method is that we can increase the learning rate of the network without fear of excessive weight gain, lack of convergence, and instability of the network.

## III. EXPERIMENTS AND RESULTS

### A. Data and Experimental Setup

The FARSDAT dataset collected by Bijankhan et al. contains a continuous and clean spoken signal from 304 male and female speakers that are different in age, dialect, and level of education [31]. Each of the speakers expresses 20 sentences in two parts. These databases are cut off and labeled at the phoneme level with windows of 23 milliseconds by step forward of half-length of these windows. FARSDAT Database Labeling is done by people familiar with linguistics and with the help of related software [31]. We used 297 speakers to train the networks and 7 speakers to test them. Various methods, including Mel-frequency Cepstral Coefficients (MFCC), Perceptual Linear Prediction Coefficients (PLP), and Logarithm of Hamming Critical Filter Bank Coefficients (LHCB), are available to extract the speech signal features. According to the results presented by Rahiminejad in 2003, extracting features of the FARSDAT dataset using LHCB parameters is better than other methods [32], therefore, this method has been used to extract the features of spoken signals. Consequent to this, the frequency spectrum of the speech signal should be partitioning into contexts each of which identifies a phoneme between 30 phonemes of the Farsi language. The length of the contexts in most experiments carried out in this work are 15 or 18 and in special cases 12. The initial learning rate at the beginning of the training was set to 0.1 and after each epoch, the performance of the network was evaluated on the test data. If the percentage of accuracy obtained from this epoch was less than previous epoch, the learning rate would be divided into 2, the weights obtained in this stage were eliminated and used from previous epoch weights to start the training. If this happens 5 times, training of the network stops automatically and the results are saved. We have used a framework that published by Palm for implementations [33].

### B. Implementations on FARSDAT dataset

In this section, first, we will focus on fully connected networks and then deal with CNNs.

#### 1) Fully connected networks

In this part, implementations will be described on all types of fully connected networks with neurons with different operator functions. In order to obtain the best and most efficient performance of the neural networks, the depth of the network and its number of parameters must be determined in proportion to the size and dimensions of the data. For this purpose, various experiments were carried out on network structures with different depths and volumes, so that we could compare the results with the best model. For the structures with sigmoid neurons, the layer-by-layer pre-training method is used to train them. According to Table 1, the number of layers indicates that the depth required for the network to have the best performance on the FARSDAT database is only three layers. All percentages are based on frame recognition rate. This post confirms the report presented in 2014 by Jimmy Ba et al. They demonstrated empirically that, shallow neural networks can learn complex functions that were previously only learnable by deep networks [34]. Of course, adding more training data will likely require more depth in the network structure. Another point that concerns the Maxout networks is their convergence with the number of epochs less than sigmoid and ReLU units. By comparing the number of convergence epochs in Table 1, we find that Maxout networks converge with or less than half the number of convergence epoch for other models. This property can be attributed to the high flexibility that this neuronal model presents against the various inputs. The reason is that the function of this neuron determines its input and output relationship during the training process and based on input data in contrast to the sigmoid and ReLU models that have a constant relationship from the beginning to the end of the training.

*Table 1. Performance Comparison between Different Structures of fully connected networks with Sigmoid, ReLU, and Maxout Neurons*

| Structure | Properties | | | |
|---|---|---|---|---|
| | Neuron Model | Epoch | Time | Acc |
| **FC**-2hidden-400 | Sigmoid | 18 | 48 min | 83 |
| **FC**-3hidden-400 | Sigmoid | 18 | 75 min | 82.78 |
| **FC**-5hidden-400 | Sigmoid | 25 | 3 h | 82.76 |
| **Pre-trained_FC** 500-500-400-400-400 | Sigmoid | 51 | 17 h | 85.21 |
| **FC**-2hidden-400 | ReLU | 15 | 51 min | 86.38 |
| **FC**-4hidden-400 | ReLU | 12 | 100 min | 88.13 |
| **FC**-2hidden-200 | Maxout | 15 | 60 h | 87.47 |
| **FC**-2hidden-400 | Maxout | 17 | 94 h | 88.14 |
| **FC**-3hidden-400 | Maxout | 16 | 4 h | 88.34 |
| **FC**-5hidden-400 | Maxout | 14 | 6 h | 88.08 |
| **FC**-6hidden-400 | Maxout | 12 | 6.5 h | 87.94 |

By comparing the results of experiments performed on fully connected networks with different neuronal models, we will notify the proper functioning of the networks with ReLU and especially Maxout models. With the increase in the number of layers and the depth of the Maxout networks, they are easily able to pass through the local minima and reach the optimal point. But networks with sigmoid neuronal models, as long as the depth of the network increases, have reduced performance (without pre-training), greatly due to the presence of local minima. The performance of sigmoid, ReLU, and Maxout neuronal models are compared in Fig. 2. In each case, the best structure obtained from several experiments associated with each neuronal model was used and for the sigmoid neuron, a layered by layer pre-training method was used to cross the local minima. ReLU models converge with the number of epochs less than the pre-trained sigmoid neuron and result in a higher recognition percentage. The same is true for the Maxout models in comparison to the ReLU models.

*2) Convolutional Neural Networks*

In this section, the performance of CNNs is evaluated. Considering that in the previous section we concluded that the Maxout neuronal model is better than other models, all neurons in the structure of networks have been selected to be Maxout. The weight-sharing process along any given dimension will have the ability to model small spatial variations in that dimension. In most prior works, one-dimensional convolutional structures along the frequency side have been used for speech processing and recognition applications [15][1][35][23][36][37][14]. To evaluate the performance of convolutional networks, experiments were carried out on one-dimensional structures, some of which were performed along the dimension of time, and some others along the dimension of frequency. In all experiments, we use a weight limitation for network stability. A two-dimensional CNN-based structure was also used to compare its performance with a one-dimensional network. According to the results reported in Table 2, we can say that weight sharing along both axes with two-dimensional CNN has a better performance than sharing along the frequency or time dimension alone. But if more optimal methods are used to combine time and frequency information of the speech spectrum signal, the performance of the speech recognition system will certainly be improved. Various Dropout rates were used and P = 0.7 was their most effective. When we used Dropout, the number of training epochs was increased. Of course, these results are due to the preprocessing performed by the process of extracting LHCB parameters to reduce the FARSDAT dimension and may be different from other methods such as MFCC or PLP. In 2015, Palaz et al. showed that CNNs have a better ability to model phonetic classes by raw signal than other extraction-based methods [38]. For this reason, it can be said that the same function of convolutional and fully connected networks is due to the use of LHCB parameters to reduce the data dimension or the power of generalization and optimization of Maxout and ReLU neurons. However, convolutional networks converge with a lower number of epochs.

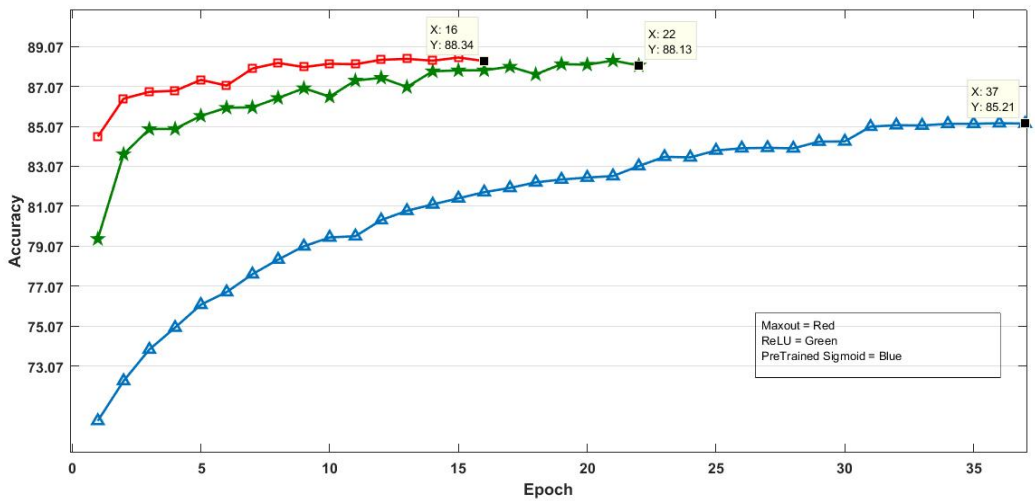

*Figure 2. Performance Comparison of fully connected networks with Sigmoid, ReLU, and Maxout neuronal models.*

*Table 2. Performance evaluation of different weight sharing methods in Convolutional Neural Networks (T stands for time, F stands for frequency and D represents the Dropout rate).*

| Structure | Properties ||||| 
|---|---|---|---|---|---|
| | Weight Sharing | D | Epoch | Training Time | Acc |
| **1D-CMNN** C80_K5_S2_F600 | F | – | 8 | 2 Days | 87.94 |
| **1D-CMNN** C40_K3_S2_C40_K3_S2_F600 | T | – | 16 | 3 Days | 88.08 |
| **2D-CMNN** C80_K7_S2_F400_F400 | T & F | – | 8 | 3 Days | 88.57 |
| **2D-CMNN** C40_K7_S2_F400_F400 | T & F | 0.3 | 15 | 3 Days | 87.97 |
| **2D-CMNN** C40_K7_S2_F400_F400 | T & F | 0.5 | 12 | 2 Days | 88.88 |
| **2D-CMNN** C40_K7_S2_F400_F400 | T & F | 0.7 | 12 | 3 Days | 89.42 |

IV. CONCLUSION

The purpose of this work was to evaluate and compare the performance of various structures of deep neural networks in continuous speech recognition. Fully connected neural networks (FCNNs) and later Deep Neural Networks (DNNs) were among the first models used in speech recognition devices. But these structures had weaknesses in their structure to model the main nature of speech signals. To overcome the weaknesses of these networks, new methods and structures including neural network pre-training methods, Convolution Neural Networks, Dropout, appropriate neuronal models such as ReLU, and Maxout, were presented. The details and method of operation of these structures and methods are described. To find the optimal structure, several experiments were implemented and the most important results were presented. After verifying the efficiency of these methods, the structures were implemented on FARSDAT Persian speech dataset. According to the results of the experiments, the combined model CMDNN (Convolutional Maxout Deep Neural Network) has improved the recognition performance of DNNs on the FARSDAT dataset to nearly 3% compared to the pre-trained fully connected structure with sigmoid neurons.

REFERENCES

ADDIN Mendeley Bibliography CSL_BIBLIOGRAPHY [1]
G. Hinton *et al.*, "Deep neural networks for acoustic modeling in speech recognition: The shared views of four research groups," *IEEE Signal Process. Mag.*, vol. 29, no. 6, pp. 82–97, 2012.

[2] M. Cai and J. Liu, "Maxout neurons for deep convolutional and LSTM neural networks in speech recognition," *Speech Commun.*, vol. 77, pp. 53–64, 2016.

[3] A. Waibel, "Modular construction of time-delay neural networks for speech recognition," *Neural Comput.*, vol. 1, no. 1, pp. 39–46, 1989.

[4] S. Hochreiter and J. J. Schmidhuber, "Long short-term memory," *Neural Comput.*, vol. 9, no. 8, pp. 1–32, 1997.

[5] Y. LeCun, L. Bottou, Y. Bengio, and P. Haffner, "Gradient-based learning applied to document recognition," *Proc. IEEE*, vol. 86, no. 11, pp. 2278–2324, 1998.

[6] S. A. Z. Seyyedsalehi and S. A. Z. Seyyedsalehi, "A fast and efficient pre-training method based on layer-by-layer maximum discrimination for deep neural networks," *Neurocomputing*, vol. 168, pp. 669–680, 2015.

[7] G. E. Hinton and R. R. Salakhutdinov, "Reducing the dimensionality of data with neural networks," *Science (80-. ).*, vol. 313, no. 5786, pp. 504–507, 2006.

[8] X. Glorot, A. Bordes, and Y. Bengio, "Deep sparse rectifier neural networks," in *Proceedings of the Fourteenth International Conference on Artificial Intelligence and Statistics*, 2011, pp. 315–323.

[9] I. J. Goodfellow, D. Warde-Farley, M. Mirza, A. Courville, and Y. Bengio, "Maxout networks," *arXiv Prepr. arXiv1302.4389*, 2013.

[10] S. A. Z. Seyyedsalehi, "Bidirectional Layer-By-Layer Pre-Training Method for Deep Neural Networks Training (In Persian)," *Comput. Intell. Electr. Eng.*, pp. 1–10, 1394.

[11] Y. LeCun and Y. Bengio, "Convolutional networks for images, speech, and time series," *Handb. brain theory neural networks*, vol. 3361, no. April 2016, pp. 255–258, 1995.

[12] O. Abdel-Hamid, A. Mohamed, H. Jiang, and G. Penn, "Applying convolutional neural networks concepts to hybrid NN-HMM model for speech recognition," in *Acoustics, Speech and Signal Processing (ICASSP), 2012 IEEE International Conference on*, 2012, pp. 4277–4280.

[13] H. Lee, P. Pham, Y. Largman, and A. Y. Ng, "Unsupervised feature learning for audio classification using convolutional deep belief networks," in *Advances in neural information processing systems*, 2009, pp. 1096–1104.

[14] C. Paper, S. Processing, and I. Conference, "Applying Convolutional Neural Networks concepts to hybrid NN-HMM model for speech recognition," no. July 2015, 2012.

[15] H. J. L. D. G. P. D. Y. Ossama Abdel-Hamid Abdel-rahman Mohamed, "Convolutional Neural Networks for Speech Recognition," *IEEE/ACM Trans. Audio, Speech, Lang. Process.*, vol. 22, no. 10, pp. 1533–1545, 2014.

[16] T. N. Sainath *et al.*, "Improvements to deep convolutional neural networks for LVCSR," 2013.

[17] P. Dayan and L. Abbott, "Theoretical Neuroscience: Computational and Mathematical Modeling of Neural Systems (Computational Neuroscience)," *J. Cogn. Neurosci.*, p. 480, 2002.

[18] M. D. Zeiler *et al.*, "ON RECTIFIED LINEAR UNITS FOR SPEECH PROCESSING New York University, USA Google Inc ., USA University of Toronto, Canada," *New York*, pp. 3517–3521, 2013.

[19] G. E. Dahl, T. N. Sainath, and G. E. Hinton, "Improving deep neural networks for LVCSR using rectified linear units and dropout," in *Acoustics, Speech and Signal Processing (ICASSP), 2013 IEEE International Conference on*, 2013, pp. 8609–8613.

[20] A. L. Maas, A. Y. Hannun, and A. Y. Ng, "Rectifier Nonlinearities Improve Neural Network Acoustic Models," *Proc. 30 th Int. Conf. Mach. Learn.*, vol. 28, p. 6, 2013.

[21] L. Tóth, "Phone recognition with deep sparse rectifier neural networks," in *Acoustics, Speech and Signal Processing (ICASSP), 2013 IEEE International Conference on*, 2013, pp. 6985–6989.

[22] L. Tóth, "Convolutional deep rectifier neural nets for phone recognition.," in *Interspeech*, 2013, no. August, pp. 1722–1726.

[23] M. Cai, Y. Shi, J. Kang, J. Liu, and T. Su, "Convolutional maxout neural networks for low-resource speech recognition," in *Chinese Spoken Language Processing (ISCSLP), 2014 9th International Symposium on*, 2014, pp. 133–137.

[24] Y. Miao, F. Metze, and S. Rawat, "Deep maxout networks for low-resource speech recognition," in *Automatic Speech Recognition and Understanding (ASRU), 2013 IEEE Workshop on*, 2013, pp. 398–403.

[25] M. Cai, Y. Shi, and J. Liu, "Deep maxout neural networks for speech recognition," in *Automatic Speech Recognition and Understanding (ASRU), 2013 IEEE Workshop on*, 2013, pp. 291–296.

[26] Y. Miao and F. Metze, "Improving language-universal feature extraction with deep maxout and convolutional neural networks," 2014.

[27] P. Swietojanski, O. M. Way, J. Li, and J.-T. Huang, "Investigation of maxout networks for speech recognition," in *Acoustics, Speech, and Signal Processing (ICASSP), 2014 IEEE International Conference on*, 2014, pp. 7649–7653.

[28] Y. Zhang, M. Pezeshki, P. P. Brakel, S. Zhang, C. L. Y. Bengio, and A. Courville, "Towards end-to-end speech recognition with deep convolutional neural networks," *arXiv Prepr. arXiv1701.02720*, 2017.

[29] N. Srivastava, G. Hinton, A. Krizhevsky, I. Sutskever, and R. Salakhutdinov, "Dropout: A Simple Way to Prevent Neural Networks from Overfitting," *J. Mach. Learn. Res.*, vol. 15, pp. 1929–1958, 2014.

[30] N. Srebro, J. D. M. Rennie, and T. S. Jaakkola, "Maximum-Margin Matrix Factorization," *Adv. Neural Inf. Process. Syst.*, vol. 17, pp. 1329–1336, 2005.

[31] M. Bijankhan, J. Sheikhzadegan, and M. R. Roohani, "FARSDAT-The speech database of Farsi spoken language," 1994.

[32] S. A. S. Mahdi Rahiminejad, "A Comparative Study of Representation Parameters Extraction and Normalization Methods for Speaker Independent Recognition of Speech (In Persian)," *Amirkabir*, vol. 55, p. 20, 1382.

[33] R. B. Palm, "Prediction as a candidate for learning deep hierarchical models of data," *Tech. Univ. Denmark*, vol. 5, pp. 1–87, 2012.

[34] L. J. Ba, R. Caruana, J. Ba, and R. Caruana, "Do deep nets really need to be deep?," in *Advances in neural information processing systems*, 2014, pp. 2654–2662.

[35] O. Abdel-Hamid, L. Deng, and D. Yu, "Exploring Convolutional Neural Network Structures and Optimization Techniques for Speech Recognition," *14th Annu. Conf. Int. Speech Commun. Assoc. (INTERSPEECH 2013)*, no. August, pp. 3366–3370, 2013.

[36] L. Tóth, "Convolutional deep maxout networks for phone recognition," *Proc. Annu. Conf. Int. Speech Commun. Assoc. INTERSPEECH*, no. September, pp. 1078–1082, 2014.

[37] L. Deng, O. Abdel-Hamid, and D. Yu, "A deep convolutional neural network using heterogeneous pooling for trading acoustic invariance with phonetic confusion," in *Acoustics, Speech, and Signal Processing (ICASSP), 2013 IEEE International Conference on*, 2013, pp. 6669–6673.

[38] M. Magimai, D. Palaz, M. M.- Doss, and R. Collobert, "Convolutional neural networks-based continuous speech recognition using raw speech signal," in *Acoustics, Speech, and Signal Processing (ICASSP), 2015 IEEE International Conference*

*on*, 2015, no. November, pp. 4295–4299.